# Estimation of Missing Data Using Computational Intelligence and Decision Trees

George Ssali and Tshilidzi Marwala
School of Electrical and Information Engineering
University of the Witwatersrand
Private Bag x3
Wits, 2050
South Africa
E-mail: tshilidzi.marwala@wits.ac.za

Abstract
This paper introduces a novel paradigm to impute missing data that combines a decision tree with an auto-associative neural network (AANN) based model and a principal component analysis-neural network (PCA-NN) based model. For each model, the decision tree is used to predict search bounds for a genetic algorithm that minimize an error function derived from the respective model. The models' ability to impute missing data is tested and compared using HIV sero-prevalance data. Results indicate an average increase in accuracy of 13% with the AANN based model's average accuracy increasing from 75.8% to 86.3% while that of the PCA-NN based model increasing from 66.1% to 81.6%.

Key words: missing data, decision trees, neural networks

## 1 Introduction

Missing data is a widely recognized problem affecting large databases that creates problems in many applications that depend on access to complete data records such as data visualization and reporting tools. This problem also limits data analysts interested in making policy decisions based on statistical inference from the data and thus imputing missing data is often invaluable as it preserves information and produces better, less biased estimates than simple techniques (Fogarty, 2006) such as list-wise deletion and mean-value substitution.

To illustrate the missing data problem, a real-world database consisting of HIV sero-prevalance data is used that was collected from an antenatal clinic survey conducted in South Africa in 2001. Experimentation using some novel data-imputation techniques has been performed using this data: one technique provides a significant improvement in the accuracy of previous work performed using Auto-Associative Neural Network and the Genetic Algorithm (Abdella and Marwala, 2005) by combining this architecture with a decision-tree based machine learning algorithm. The other approach combines a Principal Component Analysis (PCA) model with a Neural Network and uses the Genetic Algorithm, along with a decision-tree, to impute missing data.

## 2 The Missing Data Problem

The missing data problem is common and often unavoidable especially when dealing with large data sets from several real-world sources. In the context of missing data in surveys, the problem has been studied extensively (Huisman, 2000; Schafer and Graham, 2002; Little and Rubin, 2002) and can arise from non-response by the interviewee or poorly designed questionnaires. Industrial databases also experience this problem especially in cases where important information (for example information on the equipment's operational environment) has to be entered manually (Lakshminarayan et al., 1999) or due to instrumentation failures. This section presents a brief overview of existing methods for dealing with missing data based on the missing data mechanism defined by Little and Rubin (Little and Rubin, 2002).

*2.1 Missing Data Mechanism*

The missing data mechanism is usually classified as (Little and Rubin, 2002):
- Missing Completely At Random (MCAR) whereby the probability of a missing value for a variable is unrelated to the variable's value itself or to any other variable in the data set.
- Missing At Random (MAR) which arises if the probability of missing data of a particular variable could depend on other variables in the data set but not on the variable's value itself.
- Non ignorable case whereby the probability of missing data is related to the value of the variable even if other variables in the analysis are controlled.

The approach used to handle missing data depends on the situation and the mechanism listed above. A quick overview of the current techniques is presented in the next section.

*2.2 Omission of Incomplete Records*

This strategy is applicable when the number of incomplete records is small in comparison to that of complete records, simply ignores or removes records with any missing values. Furthermore, this approach must only be performed if analysis of the remaining complete data set will not lead to biased estimates (Allison, 2002) and it is rather wasteful since it usually decreases the information content of the data.

*2.3 Missing Data Imputation*

Here missing values are estimated based on those values that are available. Imputation techniques can be split into procedures based on non-model based and model-based approaches (Lakshminarayan et al., 1999). Common non-model based procedures include:
- Mean Imputation where the missing value is replaced with the mean of all reported values for that attribute.
- Hot-deck imputation where the missing value is replaced with a value from another similar case for which that value is available.

Whilst these methods are easy to use and preserve the data, they tend to attenuate variance estimates in statistical procedures and furthermore, because substituted means are not independent from other observations in the data, analyses using mean imputation have less degrees of freedom than is warranted (Hawkins and Merriam, 1991).

Model-based imputation on the other hand is more flexible and less ad hoc than the above procedures (Fogarty, 2006). These commonly include:

- Regression Techniques that estimate a missing value using a regression equation-based model derived from previously observed complete cases.
- Likelihood Based Approaches like Expectation Maximization (EM) (Dempster et al., 1977) and variants of it such as the raw maximum likelihood (also known as Full Information Maximum Likelihood (FIML) that fit probabilistic statistical models of the data. The EM algorithm uses the following strategy: first, impute the missing data values; secondly, estimate the data model parameters using these imputed values; next, re-estimate the missing data values using these estimated model parameters and repeat, iterating until convergence (Pearson, 2006).

The above single imputation strategies possess a disadvantage in that they tend to artificially reduce the variability in the estimated data. This provides the motivation for using multiple imputation techniques where several imputed data sets are generated and subjected to the same analysis to give a set of results from which variability estimates (e.g. standard deviation) and other typical characterizations (e.g. mean) can be computed (Pearson, 2006). In this paper we focus on the imputation of missing data, concentrating mainly on continuous variables, by combining model-based and non-model based procedures as discussed below.

**3 Decision-Trees, Neural Network and Genetic Algorithm Approach**

This section describes the combination of two machine-learning systems: a decision-tree based classifier algorithm and back-propagation for neural networks, along with the genetic algorithm optimization routine to impute missing data.

*3.1 Decision Trees*

A decision tree is basically a classifier that shows all possible outcomes and the paths leading to those outcomes in the form of a tree structure. Various algorithms for inducing a decision tree are described in existing literature for example CART (Classification and Regression Trees) (Breiman et al., 1984), OC1 (Murthy et al., 1993), ID3 and C4.5 (Quinlan, 1993). These algorithms build a decision tree recursively by partitioning the training data set into successively purer subsets. A high-level description of the basic algorithm is as follows (Salzberg, 1995; Tan et al., 2005):

For a given set of training records, *St*, associated with a node t, let *$C_i$ for i = {1, 2, ...,m}* be the class labels.
1. Split *St* into smaller subsets using a test on one or more attributes.
2. Check the split results. If all subsets are pure (all records in St belong to the same class *Ct*), label the leaf node with the class name *Ct* and stop.
3. Recursively split any partitions that are not pure.

The splitting criterion typically determines the difference between trees created with different algorithms. The C4.5 algorithm used in this investigation measures the entropy of the initial set and subsets produced after splitting, and chooses attributes with the most information gain based on the Information theory (Quinlan, 1993; Han and Kamber, 2000). So if *S* contains *si* tuples of class *Ci* then the Information (entropy) required to classify any given tuple is given by (Han and Kamber, 2000):

$$I(s_1, s_2, \ldots, s_m) = -\sum_{i=1}^{m} \frac{s_i}{s} \log_2\left(\frac{s_i}{2}\right) \qquad (1)$$

Assuming that an attribute *A* with *v* values is selected as a candidate root of a given tree, and then *S* will be partitioned into sets *{S1, S2, ...Sv}*. The expected information needed to complete the tree with *A* as the root is:

$$E(A) = \sum_{j=1}^{v} \frac{s_{1i} + \ldots + s_{mj}}{s} I(s_{1j}, \ldots, s_{mj}) \qquad (2)$$

And thus the information gained by branching on attribute A is given by:

$$Gain(A) = I(s_1, s_2, \ldots, s_m) - E(A) \qquad (3)$$

Under C4.5, the attribute with the highest information gain is chosen to branch the given tree. A more detailed explanation on how C4.5 builds and prunes decision trees can be found in (Quinlan, 1993).

*3.2 Neural Network and Genetic Algorithm for Missing Data*

Neural networks (NNs) are a modeling technique based on observed behavior of biological neurons. They consist of a number of interconnected nodes (neurons) tied together with weighted connections whose values are modified by a learning algorithm to yield a mapping between input and output data sets (Abdi, 1994). A typical learning algorithm, back-propagation, compares the NN's output to the desired output, calculates an error, propagates it back through the network and using a non-linear optimizing algorithm such as gradient descent adjusts the NN's weights to decrease this error (Rumelhart et al., 1986). An Auto-Associative Neural Network (AANN) is a network that is trained to predict its input vector, that is, the network's input neurons are also its output neurons.

The Genetic Algorithm (GA) optimization routine is a "survival of the fittest" evolution theory-inspired algorithm that moves from one population of possible solutions (chromosomes) to another, using a form of "natural selection" together with genetics inspired operators of crossover and mutation. The "fitness" of a given solution is determined by a "fitness function" that assigns a score (fitness) to the solution based on how well it solves the problem at hand. A selection operator selects chromosomes in a given population allowed to reproduce and on average the fitter chromosomes produce more offspring than the less fit ones(Mitchell, 1999).

A technique that combines AANNs with the Genetic Algorithm optimization routine was used by Abdella and Marwala to impute missing data in an industrial database (Abdella and Marwala, 2005). For a well chosen AANN autoencoder architecture, one expects the inputs to be very similar to the outputs. In reality however, the input and output vectors are not the same thus giving an error expressed as:

$$\varepsilon = \vec{X} - f(\vec{X}, \vec{W}) \tag{4}$$

Where $\vec{X}$ and $\vec{W}$ are the input and weight vectors respectively. Suppose that $\vec{X}$ contains $X_k$ known and $X_u$ unknown entries and the error function is squared to ensure that it is always positive then Equation 4 can be written as:

$$\varepsilon = \left( \begin{pmatrix} x_k \\ x_u \end{pmatrix} - f\left( \begin{pmatrix} x_k \\ x_u \end{pmatrix}, \vec{W} \right) \right)^2 \tag{5}$$

The GA optimization routine is then used to guess missing values that minimize the error function in Equation 5.

*3.3 Combining C4.5 with AANN-GA*

The AANN-GA technique is a model-based data imputation technique and these model-based techniques are better suited for users familiar with the missing data problem and possess the necessary expertise to apply their knowledge in building an accurate model. C4.5 on the other hand is not model-based as it does not make any assumptions on the data parameters. It has been used for data completion by treating missing value imputation of discrete valued attributes as a classification task (Lakshminarayan et al., 1999) but the database in this case was largely made up of continuous attributes which C4.5 does not naturally handle.

The combination of these two procedures is shown in Figure 1. C4.5 is used to classify intervals of the missing continuous attributes and these intervals are then used as the bounds in which the GA searches for the missing value. GA bounds are commonly set to the entire normalized range for the attribute (0 to 1 say) but empirical results show that by limiting the GA bounds using this approach, there is a significant improvement in the accuracy of the AANN-GA architecture as will be shown later.

**4 Decision Tree, PCA-NN-GA Approach**

This section describes the use of a Principal Component Analysis (PCA) model with a Neural Network (NN) and Genetic Algorithm to impute missing data.

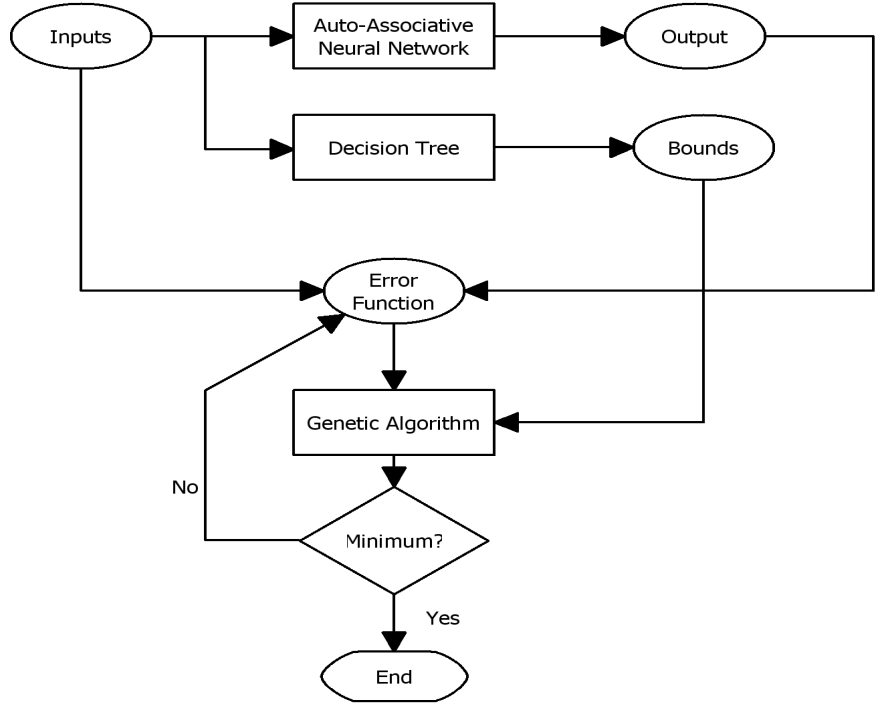

FIGURE 1. Decision Tree, Auto-associative Neural Network and Genetic Algorithm

*4.1 Principal Component Analysis*

Principal Component Analysis is a popular statistical technique commonly used to find patterns in high-dimensional data and to reduce these dimensions (Jollife, 1986). It has been applied in several fields such as face recognition (Turk and Pentland, 1991), enhancing visualization of high-dimensional data (Wolfgang et al., 2006), image compression (Ye et al., 2004), to mention but a few.

By identifying the major causes of variation in a data set, a PCA analysis provides a more compressed description of the data. Furthermore the principal components, ordered by importance in accounting for as much of the variation in the data as possible, provide a basis for dimension reduction by selecting the major principal components and omitting the less significant ones. An overview of the PCA notation is now presented as follows:

Let $X_{M \times N}$ be the input set of M records with N dimensions after centralization (i.e. after subtracting the sample mean from each input). The aim is to derive a mapping $\psi: \chi \mapsto Z$ that maps the input features into a K-dimensional space with K < N. The mapping is given by:

$$Y_{K \times M} = (U_{N \times K})^T \cdot X_{M \times N}^T \qquad (6)$$

where $U_{N \times K}$ is a feature vector made up of $K$ principal eigenvectors of the covariance matrix of $X$.

*4.2 Proposed Data-Imputation Model*

The proposed PCA-NN-GA imputation model is shown in Figure 3.

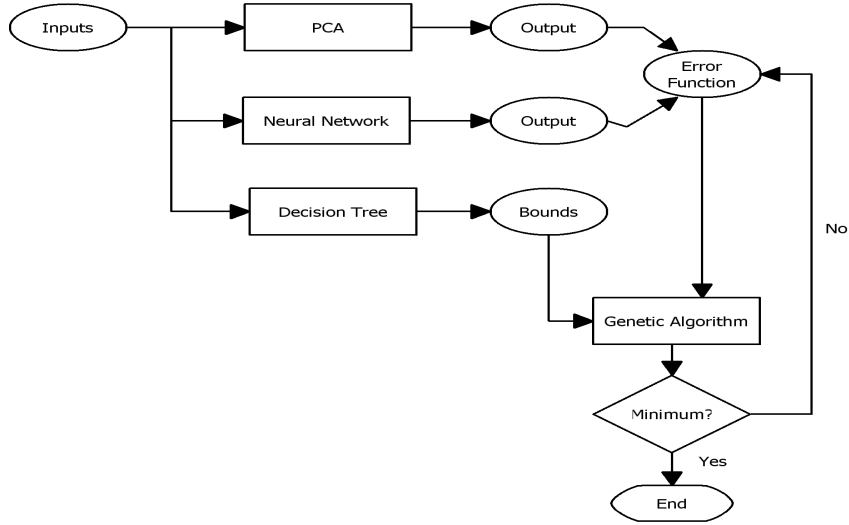

**FIGURE 2: Proposed PCA-NN-GA data imputation model**

An MLP neural network is trained to map the input set $X_{M \times N}$ to the transpose of the reduced-dimension data set in Equation 6, thereby creating two models (PCA and NN) for the data that should produce identical outputs. The difference between the outputs of these two models is used as the error function for the GA algorithm to minimize when searching for a missing value. Thus Equation 5 is modified to become:

$$\varepsilon = \left( f_{PCA} \begin{Bmatrix} x_k \\ x_u \end{Bmatrix} - f_{NN} \left( \begin{Bmatrix} x_k \\ x_u \end{Bmatrix}, \vec{W} \right) \right)^2 \qquad (7)$$

Where $f_{PCA}$ is the PCA model's function that transforms the input vector $\vec{X}$ using the feature vector of $K$ selected eigenvectors, $\vec{U}_K$. $f_{NN}$ is the Neural Network model's function and the rest of the notation is the same as before (see Equation 5). Once again, the C4.5 decision tree algorithm is applied to this architecture to predict GA search bounds for the various missing variables.

## 5 Experiments and Results

This section presents the experiments carried out using the aforementioned procedures to predict missing values in the HIV data set. To simplify the procedure, it was assumed that the data was missing at random and the missing data mechanism was ignorable (Schafer, 1997).

*5.1 Experimental Data*

The experimental data used was from an antenatal clinic survey conducted in South Africa in 2001. It consisted of the following attributes:
- Age: ranging from 14 to 50 years.

- Education Level: where a number ranging from 0 to 13 represents the highest school grade completed by a candidate with 13 indicating tertiary level education.
- Father's Age: the age of the father responsible for the most recent pregnancy.
- Gravidity: the number of times a candidate has fallen pregnant.
- Parity: the number of successful pregnancies.
- Race: encoded in a binary fashion to cover five possible options which are Asian, Black, Colored, Other and White.
- Province: encoded in a binary fashion to cover all 9 provinces in South Africa.
- HIV status of the candidate with a 0 or 1 representing negative or positive status respectively.
- 

All records with missing field were removed, the most prevalent being education level with 22% missing, as well as outliers and records with logical errors leaving a total of 12179 records from an initial 16743 records. These where then randomly mixed and split to form a training data set (9745 records), validation set for use in early-stopping explained shortly (1217 records) and a testing data set consisting of previously unseen data with missing values for various attributes to be predicted (1217 records).

*5.2 Experiment 1*

A Multi-Layer Perceptron (MLP) neural network consisting of 13 input nodes for the aforementioned variables, 11 hidden nodes and 13 output nodes was used to build a model of the data (Nabney, 2007; MathWorks, 2007). Training was performed using the Scaled Conjugate Gradient (SCG) supervised learning algorithm (Moller, 1993). The number of hidden nodes was chosen experimentally, for a fixed number of training cycles, to minimize the Root Mean Square Error (RMSE) over the training data set: Where $y$ and $y_0$ are the true and actual network outputs respectively and $n$ is the size of the data set. In order to minimize redundancy in the autoencoder, the investigated number of hidden nodes was always less than the number of input nodes (Betechuoh and Marwala, 2006). The number of training cycles (110) was determined in a similar fashion, for a fixed number of hidden nodes, using the early stopping method (Nelson and Illingworth, 1991) to prevent over-fitting of the data. The results of the network optimization are shown in Figure 2.

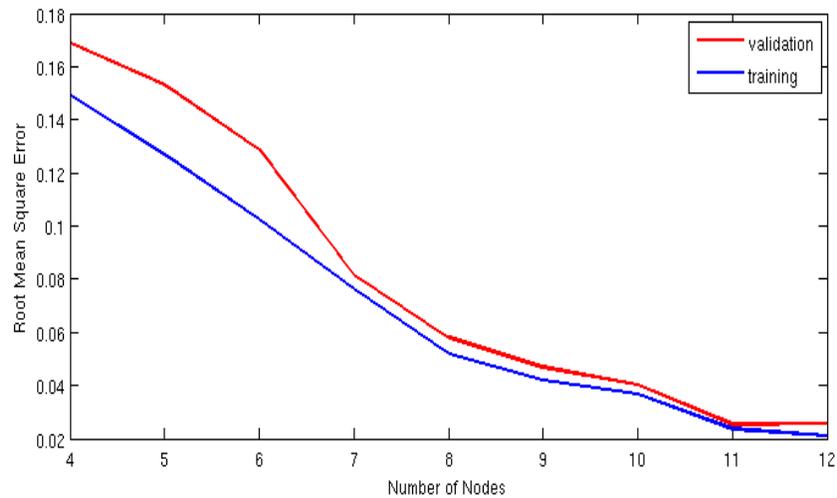
(a) Optimum Nodes

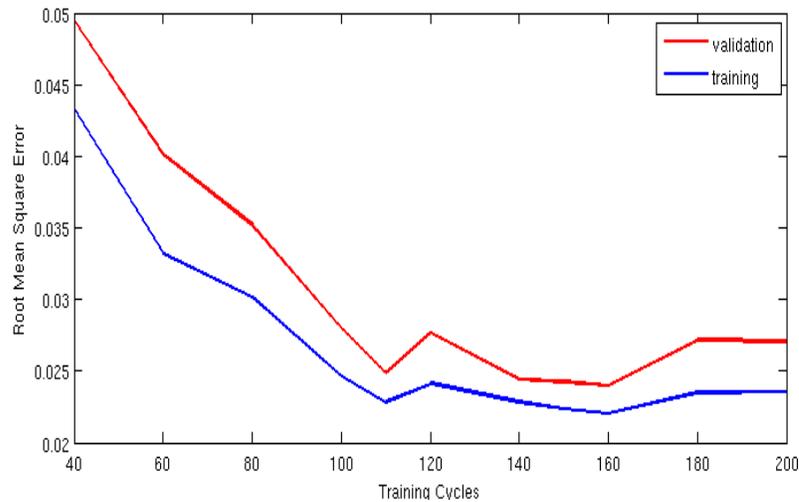
(b) Optimum Training Cycles

FIGURE 3: Auto-Associative Neural Network Optimisation Results

An implementation of the GA was conducted (Houck et al., 2007) and when predicting missing values for all attributes, the same parameters were used namely: normal geometric selection, simple crossover, non-uniform mutation, 20 generations and a population size of 50. C4.5 was trained to predict intervals for the various variables as shown in Table 1, using the training and validation data sets described in the preceding subsection. It was then applied to the test data set and the predicted intervals (bounds) passed on to the GA. Since the HIV attribute was binary and thus could be handled natively by C4.5, a C4.5 prediction of 0 translated to GA bounds of $[0 - 0.5]$ and a prediction of 1 to $[0.5 - 1]$. The experimental results presented in Table 2 above are an average from three runs of the experiment. The accuracy of imputed values was measured as a percentage of imputed values that were offset from the true value within a specified range as follows:

Table 1. Attribute intervals predicted by C4.5

| Attribute | Interval |
|---|---|
| Age | 4 years e.g 20 – 24, 25 – 29 etc. |
| Education | 2 grades e.g 0 – 2, 3 – 5, etc |
| Father's Age | 4 years e.g 20 – 24, 25 – 29, etc |
| Gravidity | 2 pregnancies e.g 0 – 2, 3 – 5, etc |
| Purity | 2 pregnancies e.g 0 – 2, 3 – 5, etc |

Table 2. Percentages of Imputed Data within the specified ranges for ANN-GA

| Method | Age | Edu | Fat | Gra | Par | HIV |
|---|---|---|---|---|---|---|
| AANN-GA | 47.7 | 32.5 | 31.4 | 80.4 | 50.9 | 77.0 |
|  | 75.0 | 46.0 | 54.7 | 97.1 | 91.0 | - |
|  | 89.0 | 59.7 | 73.0 | 99.6 | 98.5 | - |
|  | 97.0 | 76.7 | 90.8 | 100.0 | 99.7 | - |
| C4.5, AANN-GA | 52.3 | 52.1 | 41.7 | 81.8 | 60.8 | 99.7 |
|  | 79.4 | 69.5 | 68.6 | 97.8 | 92.9 | - |
|  | 89.6 | 79.4 | 82.7 | 99.7 | 98.6 | - |
|  | 97.9 | 91.8 | 93.2 | 100.0 | 99.7 | - |

- Age: to within 2, 4, 6 and 10 years.
- Education Level: to within 1, 2, 3 and 5 grades.
- Father's Age: to within 2, 4, 6 and 10 years.
- Gravidity: exact number of pregnancies and to within one, three and five pregnancies.
- Parity: exact number of pregnancies and to within one, three and five pregnancies.
- HIV status: accuracy measured using specificity.

*5.3 Experiment 2*

The PCA analysis was conducted on the combined training and validation data sets and for the different dimensions, the RMSE was calculated on the recovered data as well as the accuracy of the predicted values to within 5%. Figure 4 shows these results and as can be seen, there is a significant change in accuracy between using 6 and 7 dimensions, while using 11 dimensions only alters the accuracy slightly. Thus the new reduced dimension was chosen to be 7.

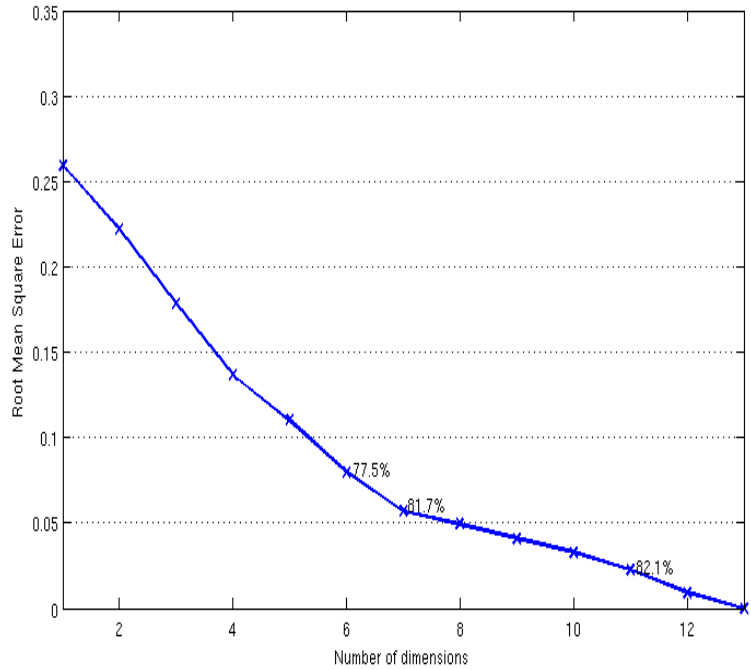

**FIGURE 4: Root Mean Square Error for varying PCA dimensions**

An MLP neural network consisting of 13 input nodes, 17 hidden nodes and 7 output nodes was set up and trained using 140 cycles using the same techniques detailed in Experiment 1 (see Section 5.2) and these optimization results are shown in Figure 5 on page 16. The GA, C4.5 and accuracy of imputed missing values was performed using the same settings and to within the same ranges as described above in Section 5.2. These results are presented in Table 3.

Table 3.Percentages of Imputed Data within the specified ranges for PCA-NN-GA

| Method | Age | Edu | Fat | Gra | Par | HIV |
|---|---|---|---|---|---|---|
| PCA-NN-GA | 30.6 | 43.5 | 20.3 | 40.6 | 38.7 | 70.1 |
|  | 55.0 | 62.6 | 37.2 | 84.1 | 72.8 | - |
|  | 71.8 | 77.2 | 50.0 | 99.0 | 91.5 | - |
|  | 91.4 | 91.0 | 69.9 | 99.7 | 98.8 | - |
| C4.5, PCA-NN-GA | 50.6 | 51.5 | 43.4 | 53.7 | 47.5 | 99.5 |
|  | 77.1 | 69.7 | 68.8 | 93.8 | 80.9 | - |
|  | 88.0 | 81.3 | 81.3 | 99.9 | 98.0 | - |
|  | 97.3 | 93.0 | 93.7 | 100 | 99.9 | - |

## 6 Discussion

*6.1 Impact of Bounds Selection*

The most glaring insight that can be gained from our experimentation is that bounds in optimization search routines matter. Whilst it can be argued that given a big enough population size and number of generations GA will find an optimum solution, this solution is not necessarily the global optimum. Furthermore, any experiment can have various parameters that can be tweaked and the experiment repeated until adequately accurate results are achieved however this is an unfair presentation of a given paradigm and it does not guarantee repeatability on another set of data. Prediction of bounds directly from given inputs, as presented in this paper, repeatedly improves the accuracy of missing data imputation irrespective of the underlying architecture of 10.5% for the AANN-GA architecture and 15.5% for the PCA-NN-GA architecture.

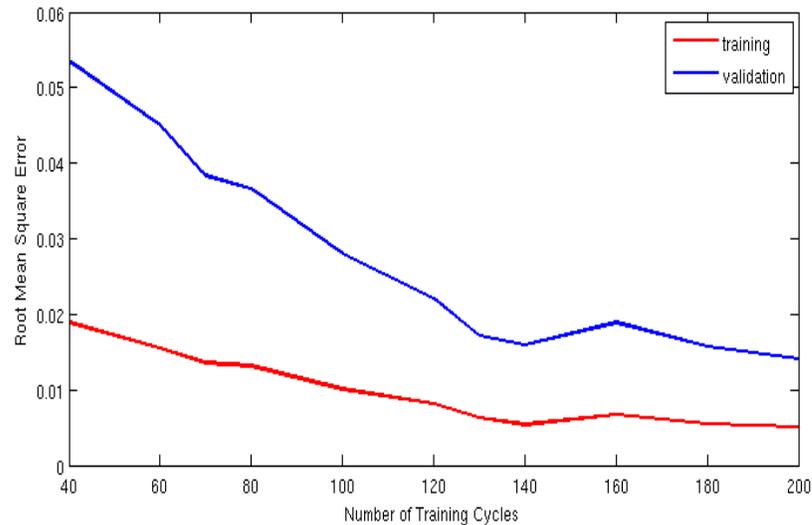

**FIGURE 5: Optimisation results for the PCA-NN-GA Neural Network**

(a) Optimum Cycles

*6.2 Note on Proposed Architectures*

Depending on the desired accuracy for a given variable, both architectures produce satisfactory results and can be used to complete incomplete databases. For example a person's age group can be accurately defined to within 4 years, 20 – 24 for early twenties and 25 – 29 for late twenties say, to which the C4.5,AANN-GA and C4.5,PCA-NN-GA architectures would give an accuracy of 79.4% and 77.1% respectively. Similarly, imputing the education variable (which had the largest number of missing values) to within 3 years accurately captures a person's level of schooling to say elementary school (grade 0 – 2), lower middle school (grade 3 – 5), upper middle school (grade 6 – 8), highschool (grade 9–11) and (highschool/college) graduate (grade 12+). The C4.5, AANN-GA predicts this education level to an accuracy of 79.4% while C4.5,PCA-NN-GA predicts to 81.3%. The Auto-Associative based architecture however performs better than the PCA based architecture for all variables except Education which could be attributed to the loss in data associated with dimensionality reduction under PCA.

**7 Conclusion**

In conclusion, a novel paradigm that combines decision trees with neural networks, principal component analysis and genetic algorithm is introduced to impute missing data. Two separate architectures, one based on an auto- associative neural network and the other principal component analysis, are set up and each is combined with a decision tree as well as the genetic algorithm optimization routine. Empirical results indicate that both architectures can adequately impute missing data and the addition of a decision tree improves results for both by an average of 13%.